\documentclass[twocolumn,preprintnumbers,aps,superscriptaddress,showpacs]{revtex4}

\def\bea{\begin{eqnarray}}
\def\eea{\end{eqnarray}}
\def\ben{\begin{equation}}
\def\een{\end{equation}}
\def\benu{\begin{enumerate}}
\def\enu{\end{enumerate}}

\def\n{n}
\def\nddot{\partial^2_t \n(\br,t)}

\def\sss{\scriptscriptstyle\rm}

\def\l{^\lambda}

\def\1var{(\bx_1...\bx\N)}

\def\br{{\bf r}}

\def\bx{{x}}

\def\bj{{\bf j}}

\def\s{_{\sss S}}
\def\xc{_{\sss XC}}
\def\Hxc{_{\sss HXC}}

\def\N{_{\sss N}}
\def\H{_{\sss H}}

\def\ext{_{\rm ext}}

\def\sph_int{ {\int d^3 r}}

\usepackage{graphicx}
\usepackage{psfrag}
\usepackage{amsmath}
\usepackage{bm}

\usepackage{epsfig}

\input epsf

\pacs{31.15.Ew,71.15.Qe}
\begin{document}
\title{Comment on ``Critique of the foundations of time-dependent density-functional theory'' [Phys. Rev A. {\bf 75}, 022513 (2007)].}
\date{\today}
\author{Neepa T. Maitra}
\affiliation{Department of Physics and Astronomy, Hunter College and City University of New York, 695 Park Avenue, New York, NY 10065, USA}
\email{nmaitra@hunter.cuny.edu}
\author{Robert van Leeuwen}
\affiliation{Department of Physics, University of Jyv\"askyl\"a, Survontie 9, FI 40014, Jyv\"askyl\"a,
Finland}
\author{Kieron Burke}
\affiliation{Department of Chemistry, 1102 Natural Sciences 2, UC Irvine, CA 92697, USA}
\begin{abstract}
A recent paper (Phys. Rev A. {\bf 75}, 022513 (2007)) challenges
exact time-dependent density functional theory (TDDFT) on several grounds.
We explain why these
criticisms are either irrelevant or incorrect, and that TDDFT is both formally
exact and predictive.
\end{abstract}
\maketitle
Time-dependent density functional theory (TDDFT) has a rigorous
foundation\cite{RG84,
L98,L99,GDP96,L01}.  Schirmer and Dreuw (SD)
appear to criticise TDDFT on several grounds~\cite{SD07},
almost all of which are
ultimately conceded by SD themselves.
For example, SD state that ``...an error is introduced in
both the TD and static KS linear response theory if the perturbing
(external) potential is given by a nonlocal operator'', but this is
misleading since such potentials do not exist in DFT.  Another
example is their long discussion of the problems of the RG action. 
Such problems were first raised a decade ago~\cite{GDP96,UEG96}, and
resolved shortly after using the Keldysh
action~\cite{L98}.  The problems with the action in Ref.~\cite{RG84}
have been thoroughly investigated in several
works~\cite{L98,L01,GDP96,MBAG02,TDDFTbook}.  
Finally, SD admit these points, but by not addressing the
current literature, they are tilting at windmills~\cite{Quixote}.  

The sole unresolved criticism of SD is
their claim that the Kohn-Sham (KS) equations of
TDDFT, even if able to {\it reproduce} the density-evolution of the
true interacting system, $\n(\br,t)$, cannot {\it predict} that evolution.
The KS potential, $v\s(\br,t)$, is defined as the unique one-body
potential in which non-interacting electrons evolve with density $\n(\br,t)$.
SD claim that their
``radical KS'' scheme shows
that $v\s(\br,t)$ functionally depends on the future density,
thereby making direct propagation impossible.
They further argue that, with neither a variational principle nor
some proof of convergence of the TD KS equations for such a
potential, TDDFT must be ``unfounded''.
The rest of our Comment addresses this issue: 
we show where the error lies in SD's arguments
and explicitly demonstrate that the KS-TDDFT procedure is indeed predictive. 

The flaw in SD's arguments arises from not
carefully distinguishing between $v\s(\br,t)$ and the
exchange-correlation (XC) potential.  If
$v(\br,t)$ is the 
time-dependent external potential, i.e., the one-body potential
applied to the interacting system (which is always known and given),
and $v\H(\br,t) = \int n(\br',t)/\vert \br - \br'\vert d^3r'$ is the Hartree potential, then
\ben
v\xc(\br,t)=v\s(\br,t)-v\H(\br,t)-v(\br,t),
\label{vs}
\een
is known to be a functional of the
initial states (both interacting, $\Psi_0$, and KS, $\Phi_0$) and
$\n(\br,t)$, written $v\xc[n,\Psi_0,\Phi_0](\br, t)$.  SD argue (pg 11)
that, since $v\s(\br,t)$ depends on $\partial^2_t n(\br,t)$ (the
second time-derivative, which SD denote as ``the future''), the
possibility of a stringent time propagation is ``thwarted'': if the
potential functional depends on the ``future'' of the density, then
how could it also possibly predict it?  In particular, at the initial
time $t=0$, $\n(\br,0), \Psi_0,$ and $\Phi_0$ are obviously
insufficient to determine $\partial^2_t n(\br,0)$, and therefore
insufficient to determine $v\s(\br,0)$.  

What SD miss is that in any
TDKS time propagation, $v(\br,t)$ is known and prescribed by the
physical problem at hand, and it is only $v\xc(\br,t)$ for which a
density-initial-state functional is needed.  Herein lies the
resolution of the propagation paradox, as the functional dependence of
$v\s(\br,t)$ generally differs from that of $v\xc(\br,t)$, in
contrast to what is claimed by SD without justification: ``A similar
temporal nonlocality must be expected for the XC potential...''.  In
particular, $v\xc[n,\Psi_0,\Phi_0](\br,0)$ depends {\it purely} on the
initial states, as we explicitly show below, and SD's expected
dependence on $\partial^2_t n(\br,t=0)$ vanishes. At other times,
unlike $v(\br,t)$ and $v\s(\br,t)$,
$v\xc(\br,t)$ does {\it not} depend on $\partial^2_t n(\br,t)$: there is 
a dependence on $\partial_t n(\br t)$  but this can be extracted from
$\Phi(t)$, through the continuity equation, and so
is available during propagation.
(There is also dependence on $\Psi_0$ and $\Phi_0$ and the
earlier history of the density).

Before demonstrating in general the density- and
initial-state-dependence of $v\xc(\br,t)$ (``potential
functionals'' in SD's notation) and why direct propagation
works, we give a simple counterexample to SD.
Propagate the time-dependent
Schr\"odinger equation for one electron in some potential from some
initial state.  Applying SD's ``radical KS'' scheme and logic, this is
impossible, because that potential functionally depends
on the future density (eg. Eq.75 of SD).  This absurd
conclusion is incorrect, because you are {\em given} $v(\br,t)$.
You never extract it from the density. Instead, what is needed for
standard KS TDDFT is a density-initial-state functional only for $v\xc(\br,t)$.
For one electron, $v\xc(\br,t)=-v\H(\br,t)$
(choosing the KS initial state as the true
initial state, the only sensible choice~\cite{MB01}), 
i.e. a functional of the instantaneous density
alone. There is no dependence on $\partial^2_t n(\br,t)$,
contrary to the ``expectations" of SD.  More generally,
because only the density-dependence of $v\xc(\br,t)$ is needed in any
TDDFT KS calculation, their result for $v\s(\br,t)$ is irrelevant
to the question of propagation.

First, to clarify the question of ``future" dependence, we note
that a dependence on $\nddot$ does not violate causality at
any finite $t$.  For any 
$0<t<T$,
$v[n;\Psi_0](\br,t)$ can depend only on the density in the
interval $[0,T]$ and not on later times because, by the RG theorem,
two potentials that are the same in $[0,T]$ but
differ for $t>T$ must give different densities at times $t>T$, i.e.,
$v$ is a causal
functional of the density. The same holds for 
$v\s[n;\Phi_0](\br,t)$.   From Eq. (1),
$v\xc[n;\Psi_0,\Phi_0](t)$ is then also a causal functional of
the density and initial states for any finite $t$, and
$\partial_t^2 n(\br,t)$ may be evaluated to the left of time t,
(e.g. using $n(t), n(t-\Delta t), n(t-2\Delta t)$).
But this argument cannot be applied at $t=0$, and so the
start of a propagation appears problematic.

To see why this is in fact not a problem, we will use results from
Ref. [5], and the notation that a superscript $(k)$ indicates the
$k$-th time-derivative at $t=0$, eg. $n^{(k)}(\br) = \partial_t^kn(\br t)\vert_{t=0}$.
Begin by noting
that both $\n(\br,0)$ and $n^{(1)} (\br)$ are determined
by $\Psi_0$, the initial wavefunction of the interacting problem,
because of continuity:
$
\partial_t \n(\br,t)= -\nabla \cdot \bj(\br,t).
$
This in turn restricts allowed choices of $\Phi_0$
to only those that recover these values.
The form of $v(\br,t)$ influences only second and
higher time-derivatives of $\n(\br,t)$.
The TD Schr\"odinger equation implies ~\cite{L99} \ben
\partial_t^2 n(\br,t) = \nabla\cdot[n(\br,t)\nabla v(\br,t)] + q(\br,t),
\label{eq:vSL}
\een
where
\ben
q(\br,t) = \langle \Psi(t)\vert\ \hat\tau(\br) +\hat w(\br)\
\vert \Psi(t)\rangle,
\label{eq:q}
\een
and
\bea
\nonumber
\hat \tau (\br)&= &\frac{1}{2} \sum_{i,k} \partial_i\partial_k\left(\partial_i\hat\psi^\dagger(\br)\partial_k\hat\psi(\br) +\partial_k\hat\psi^\dagger(\br)\partial_i\hat\psi(\br) \right. \\
\nonumber
&& -\left.\frac{1}{2}\partial_i\partial_k[\hat\psi^\dagger(\br)\hat\psi(\br)]\right)\\
\hat w(\br) &=&\sum_k\partial_k \int d^3r'\hat\psi^\dagger(\br)\hat\psi^\dagger(\br')\partial_k \frac{1}{\vert \br - \br'\vert}\hat\psi(\br')\hat\psi(\br)\,.
\nonumber
\eea
An analogous equation applies to
the KS system, with $v\s(\br t)$
and 
$q\s(\br,t) = \langle \Phi(t)\vert\ \hat\tau(\br)\
\vert \Phi(t)\rangle$.
Requiring that the density evolutions be the same yields
\ben \nabla\cdot[n(\br,t)\nabla v\Hxc (\br,t)] = q(\br,t) -q\s(\br,t)
\label{eq:central}
\een
where $v\Hxc(\br,t) = v\H(\br,t) + v\xc(\br,t)$.
This equation is of
Sturm-Liouville form: such equations have a unique solution for
$v\Hxc(\br,t)$ if $n(\br,t)$ and $q(\br,t)-q\s(\br,t)$ are given, together with
the boundary condition that $v\Hxc(r\to\infty, t) \to 0$. We shall
assume this boundary condition for all that follows; 
any choice of a TD constant for the
asymptotic potential does not affect the density[5].  Thus a KS potential
can always be found for any density for the interacting
system, provided the initial conditions are met.

We now use Eq.~(\ref{eq:central}) to show that at each time step,
the functional input to $v\xc(\br,t)$
consists of the $\Psi_0$, $\Phi_0$, 
and the density evolved through previous times. 
At $t=0$, 
\ben \nabla\cdot[n(\br,0)\nabla
v\Hxc(\br,0)] = q(\br,0) -q\s(\br,0)\,. \label{eq:0thtimestep} 
\een
Since $q(\br,0)$ and $q\s(\br,0)$ are determined by Eq.~(\ref{eq:q}), which is known entirely from the initial wavefunctions, the solution
of this equation determines $v\xc(\br,0)$. (The Hartree
potential $v\H(\br,0)$ is determined as usual directly from the
instantaneous density). Thus, as mentioned earlier, $v\xc(\br,0)$
depends on the initial states alone: the
dependence on the second-time-derivative of the density in
Eq.~(\ref{eq:vSL}) and its KS counterpart cancel once their difference
is taken in $v\Hxc[n,\Psi_0,\Phi_0](\br 0)$.  
Together with $v(\br,0)$, this
evolves $\Phi_0$ forward one time step:
\ben
\left(-\frac{1}{2}\nabla^2 + v\Hxc(\br,0) + v(\br,0)\right)\phi_i(\br,0) = i\partial_t\phi_i(\br,0) \een
This yields the
orbitals at the first time-step, $\phi_i(\br,\Delta t) = \phi_i(\br,0) + \partial_t\phi_i(\br,0)\Delta t$, from which one can obtain the
density evolved to the first time-step, $n(\br,\Delta t) =
\sum_i\vert\phi_i(\br,\Delta t)\vert^2$, as well as the current at the first time-step, ${\bf j}(\br,\Delta t) = \Im\sum_i\phi_i^*(\br, \Delta t)\nabla \phi_i(\br, \Delta t)$,
and, through the equation of continuity, also the first-time derivative of the density $\dot n(\br,\Delta t) =  -\nabla\cdot {\bf j}(\br,\Delta t)$.  

Next, consider $t=\Delta t$, for which we need to find $\partial_t v\Hxc(\br,0)$.
Take a time-derivative of Eq.~(\ref{eq:central}), evaluating the
terms at $t=0$:
\bea
\nonumber
\nabla\cdot[n(\br,0)\nabla\partial_tv\Hxc(\br,0)]  &=& -\nabla\cdot[\partial_tn(\br,0)\nabla v\Hxc(\br,0)] \\
&& + \partial_t(q(\br,0) -q\s(\br,0)) \label{eq:1sttimestep} \eea where
\bea
\nonumber
\partial_t(q(\br,0) -q\s(\br,0))&=& i\langle\Phi_0\vert[\hat \tau(\br),\hat H\s(0)]\vert \Phi_0\rangle - \\
&&i\langle\Psi_0\vert[\hat q(\br),\hat H(0)]\vert \Psi_0\rangle
\label{eq:dqdt}
\eea
Equation~(\ref{eq:1sttimestep}) is again of Sturm-Liouville form, with
a unique solution for $\partial_tv\Hxc(\br,0)$: all the other
quantities in the equation are known either from the initial states or
the results of previous time-propagation.
Eq.~(\ref{eq:1sttimestep}) is not
an explicit functional of the initial states and density alone, 
due to the appearance of 
$v(0)$ in the commutators in Eq.~(\ref{eq:dqdt}).
But 
evaluating Eq.~(\ref{eq:vSL}) at $t=0$
gives the functional dependence of $v(0)$ on
$\Psi_0, \n(\br,0),$ and $ \partial_t^2{n}(\br,0)$.
Since
$\partial_t^2{n}(0) = (\partial_t{n}(\Delta t) -
\partial_t{n}(0))/\Delta t$ (limit $\Delta t \to 0$ understood),
the appearance of the second-derivative at $t=0$ implies only a dependence
on the first derivative at $t=\Delta t$, directly available from $\Phi(\Delta t)$ via the equation of continuity.
So
\ben v\s(\br,\Delta t) =v(\br,\Delta t) + v\Hxc(\br,0) +
\partial_tv\Hxc(\br,0)\Delta t
\een
is determined, and predicts the
time-evolution of the density at $t=2\Delta$.

Each subsequent time-derivative of Eq.~(\ref{eq:central}) produces one higher 
time-derivative of the XC potential, such that $v\Hxc^{(k)}(\br)$
 is determined solely by the initial states and $n^{(l)}(\br)$ with $l\le (k+1)$, all
available from propagation to the $k$th time-step:
\ben
\nabla\cdot[n(\br, 0)\nabla v\Hxc^{(k)}(\br)] = Q^{(k)}(\br)
\label{eq:vHxc_kth}
\een
\ben
Q^{(k)}(\br) = q^{(k)}(\br) -q^{(k)}\s(\br) -\sum_{l=0}^{k-1}\left(\begin{array}{ll}k\\l\end{array}\right)\nabla\cdot[n^{(k-l)}(\br)\nabla v\Hxc^{(l)}(\br)]\;.
\label{eq:kthtime}
\een 
The $q^{(k)}(\br)$ and $q^{(k)}\s(\br)$ involve
 multiple commutators
of the operator $\hat q(\br)$ with the true and KS Hamiltonians,
respectively, and their time-derivatives, sandwiched between the {\it
initial} states $\Psi_0$, $\Phi_0$, respectively. 
For example, for $k=2$,
\bea
\nonumber
&q^{(2)}(\br)-q\s^{(2)}(\br) =&\\
\nonumber
&-\langle \Psi_0 | [\hat{q}(\br), \hat{H}(0) ] ,\hat{H}(0) ] | \Psi_0 \rangle  
+&\langle \Phi_0 | [\hat{\tau}(\br), \hat{H}\s(0) ] ,\hat{H}\s(0) ] | \Phi_0 \rangle\\
&-i\langle \Psi_0 | [\hat{q}(\br) ,d\hat{H}/dt(0) ] | \Psi_0 \rangle+&i\langle \Phi_0 | [\hat{\tau}(\br) ,d\hat{H}\s/dt(0) ] | \Phi_0 \rangle
\nonumber
\eea
The $v(0)$ and $dv(0)/dt$ appearing in the commutators
are causal density and initial-state functionals via Eq.~(\ref{eq:vSL}), so Eqs.~(\ref{eq:vHxc_kth})
and (\ref{eq:kthtime}) yield the XC potential as a causal implicit functional
of purely the density and initial states.
We reiterate that a dependence on $\partial_t \n(\br,t)$ presents
no difficulty during propagation.  As one propagates, one can evaluate this
instantaneously, by simply computing the divergence of 
the current-density of $\Phi(t)$.  Thus all quantities needed are
available from the past propagation.

So we have shown that the propagation can be done in a
predictive manner, and that the expressions for
$v\xc[n,\Psi_0,\Phi_0](\br,t)$ that
we give are causal functionals of the density and initial-state:
explicit at t=0 (Eq.~\ref{eq:0thtimestep}) and implicit at later times (via
$v[n,\Psi_0]$, Eqs.~\ref{eq:central},\ref{eq:1sttimestep},\ref{eq:dqdt},\ref{eq:kthtime}).
This construction holds for any time-dependent potentials
and densities that are equal to their Taylor expansions for $t\geq 0$
for a finite period of time (as assumed for $v(\br, t)$ in the Runge-Gross one-to-one
mapping proof). Full analyticity is not required: in particular, as
nothing is assumed for times earlier than $t=0$, this procedure
applies to sudden switch-on potentials.

The XC potentials used in practical
applications may be viewed as approximations to this formally exact
construction of the potential functional.
Although most of the applications to date have utilized adiabatic
approximations, depending on the instantaneous density alone,
memory-dependence is a well-recognized feature of time-dependent
functionals (see eg. Refs.~\cite{GK85,DBG97,MBW02,KB05, UT06, WU08}) and
memory-dependent functionals have been successfully applied to real calculations. 
There are also explicit systematic methods 
based on many-body perturbation theory to construct
approximate XC potentials for practical applications~\cite{BDLS05}.
The equations for these approximate potentials
also show the fundamental property that we demonstrated above:
the XC potentials at a given time are completely determined 
by the density evolved up to and including the present time and the initial states. 

Lastly, we give an example to show explicitly the error in SD for an interacting case.
Consider two electrons in one dimension, and we need look
only at t=0.
In one-dimension, Eq.~(\ref{eq:0thtimestep}) reduces to
\ben
v\s(x,0)= v\ext(x,0) +\int^x\frac{dx'}{n(x', 0)}\int^{x'}
dx'' (q(x,0)-q\s(x,0))
\label{eq:1dL99}
 \een
But inversion of the time-dependent KS equation\cite{HPB99,AV99}, as in SD, yields
here
\bea
\nonumber
v\s(x,0) &=& -\frac{1}{2}\left(\frac{\partial_xn}{2n}\right)^2 - \frac{1}{2}\left(\frac{j}{n}\right)^2 + \frac{\partial_x^2n}{4n} \\
&& - \int^x \left(\frac{\partial_t j}{n} + \frac{\partial_xj^2}{2n^2}\right)dx'
\label{eq:1dKSinv}
\eea
where $j = j(x,0)$ is the initial current-density, determined from the initial
wavefunction. (We drop the spatial and $t=0$ indices on the right, for ease of reading).
The alleged dependence on the future arises
through the term $\partial_t j$ on the right: this may be equivalently
written in terms of $\partial_t^2 n(x,0)$, and is the only term not directly
obtainable from the initial states.

However,
this future dependence disappears as soon
as we relate the KS system to the interacting system via the Heisenberg
equation of motion for the current of the interacting system:
\ben
\partial_t j = -n\, \partial_x v\ext - T_{xx} - W_x
\een
where $\partial_x T_{xx} = \langle \Psi_0\vert\hat\tau\vert\Psi_0\rangle$ and $\partial_x W_x = \langle \Psi_0\vert\hat w\vert\Psi_0\rangle$.
Substituting this for $\partial_t j$ in the right-hand-side of
Eq.~(\ref{eq:1dKSinv}),
\bea
\nonumber
 -\int^x \frac{\partial_t j}{n}dx' &=&\int^x \frac{dx'}{n(x',0)}\int^{x'}\partial_t^2 n(x'',0)dx'' \\
& =& v\ext(x,0) + \int^x \frac{T_{xx}+W_x}{n}dx'  
\eea
i.e. the ``future'' dependence is in the external potential and other
terms at $t=0$! That is, once the connection with the interacting system is made, 
the apparent dependence on the future evaporates as $\partial_t^2n(x,0)$ is determined
by initial-state information and by the external potential that the interacting system is subjected to. The dependence is explicit in this two-electron example, but in the general $N$-electron case, the construction of
Ref.~\cite{L99} implies that this is always true.

Inversion of the KS equations alone yields information only about
the KS potential as a functional of the density, but tells nothing about the XC potential.  How could it, since it
contains no information about any interacting system?
This is most easily
seen in the ground-state problem.  One can trivially invert the single
KS orbital equation for any two-electron density and get its KS potential,
but this tells you nothing about XC unless you know the corresponding
external potential to subtract from it.  And there's no way to find that,
without inverting the interacting Schr\"odinger equation, thereby making
the functional dependence as implicit as in the original definition~\cite{AV99,HPB99}.

Before concluding, we briefly review SD's discussion of
 the lack of a proof of numerical convergence of KS
propagation when non-adiabatic functionals are used
(the ``trajectory mode'' in SD's notation). 
SD only claim that such a proof is needed after they
incorrectly deduce that the ``potential functional" propagation
mode fails.   This is therefore a straw man argument.
There are no proofs of the existence of solutions and their convergence for,
e.g., the Navier-Stokes equations, but should they be ``abandoned?"
(In fact,
the KS propagation has been explicitly converged in a recent
calculation with a non-adiabatic potential~\cite{WU08}.)

To summarize,
(i) SD's rejection of the original RG formulation of TDDFT 
originates in an elementary logical error in their conclusions
from their ``radical KS" scheme,
(ii) this erroneous conclusion led SD
to dismiss without argument the constructive
proof of the XC potential that has long existed in the literature, in which it
is clear that the potential depends only on the past, leading to TDKS equations that are indeed predictive, 
(iii) finding a general proof of convergence for TDDFT
propagation would be interesting, but its foundations do not depend on this,
(iv) the one-to-one density-potential mapping in no way depends on
the action functional proposed in RG.  Even though 
a rigorous action principle has been proven within the Keldysh formalism~\cite{L98}, as well as in real time~\cite{V08},
it is not needed to prove that the theory is predictive.
Thus the illusion is not TDDFT, as claimed by SD, but the apparent
dependence on the future.


We are indebted to Tchavdar Todorov for invaluable discussions. 
NTM thanks the National Science Foundation's CAREER Program and the Research Corporation's Cottrell Scholar Program for financial support. KB acknowledges support from National Science Foundation grant CHE-0355405.

\end{document}